\input{aipcheck}

\documentclass[final]{aipproc}
\usepackage{amssymb}
\usepackage{amsmath}
\layoutstyle{8x11double}
\newcommand{\be}{\begin{eqnation}}
\newcommand{\ee}{\end{eqnation}}
\newcommand{\beqn}{\begin{eqnarray}}
\newcommand{\eeqn}{\end{eqnarray}}

\begin{document}

\title{Binding Energies and Melting Temperatures of Heavy Hadrons in Quark-Gluon Plasma}

\author{I.M.Narodetskii}{address={ITEP, Moscow, 117218 Russia}}
\author{Yu.A.Simonov}{address={ITEP, Moscow, 117218 Russia}}
\author{A.I.Veselov}{address={ITEP, Moscow, 117218 Russia}}

\begin{abstract}
We analyze the static potential of a quark--antiquark pair at
$T\,\geq\,T_c$, where $T_c$ is a temperature of a deconfinement
phase transition in QCD. We discuss the possibility that the
non-perturbative part of this potential can be studied through the
modification of the correlation functions, which define the
quadratic field correlators of the nonperturbative vaccuum fields.
We use the non--perturbative quark--antiquark potential derived in
this way and the screened one--gluon-exchange potential with
$T$-dependent Debye screening mass to calculate $J/\psi$,
$\Upsilon$ and $\Omega_{bbb}$ binding energies and melting
temperatures in the deconfined phase of the full 2-flavors QCD.

\end{abstract}

\maketitle


\section{Introduction}

There is a significant change of view on physical properties and
underlying dynamics of quark--gluon plasma (QGP), produced at
RHIC, see {\it e.g.} \cite{T:2009} and references there in. On
quite general grounds it is expected, that fundamental forces
between quarks and gluons get modified at finite temperature.
Instead of behaving like a gas of free quasiparticles~--~quarks
and gluons, the matter created in RHIC interacts much more
strongly than originally expected.  Recall that above the
deconfinement temperature $T_c$ the perturbative
one--gluon--exchange potential is expected to be exponentially
screened at large distances $(r\, \gg \,1/T )$ \cite{Linde:1980}.
Moreover, heavy quark free energy $F^{Q{\bar Q}}(r, T)$ shows
large asymptotic value, $F^{Q{\bar Q}}(\infty, T)$
\cite{[16]}. This value can be explained only by non-perturbative
(NP) effects, since perturbative one-gluon-exchange potential,
even with increased $\alpha_s(r)$, cannot produce similar effect.
Therefore it is more appropriate to describe the (NP)
properties of the QCD phase above $T_c$ in terms of the NP part of
the QCD force rather than a strongly coupled Coulomb force.

At $T=0$, the non-perturbative quark-antiquark potential is
$V_{np}(r)\,=\,\sigma r$, where $\sigma$ is the $SU(3)$ string
tention. This potential has been used extensively in potential
models. At $T \geq T_c$, $\sigma=0$, but that does not mean that
the NP potential disappears. Attempting to guess the form of the
non-perturbative potential we address to the Field Correlator
Method (FCM) (see \cite{NST:2009} and references there in). Within
FCM the NP $Q{\bar Q}$ potential above $T_c$ was suggested to
occur due to the non zero correlation function $D_1(x)$ that is
one of the two functions which define the quadratic field
correlators of the nonperturbative vaccuum fields . The most
direct prediction of this approach is the existence of bound
states of heavy $c{\bar c}$ and $b{\bar b}$ mesons and heavy $bbb$
baryons above $T_c$. In a recent paper \cite{NSV:2009} we
calculated binding energies for the lowest $Q\overline{Q}$ and
$QQQ$ eigenstates ($Q=c,b$) using both the NP potential of the FCM
and the screened Coulomb potential with the temperature dependent
Debay radius calculated in pure $SU(3)$ glue theory. In this talk
we discuss and extend the results of this analysis, in particular,
by including the effect of Debye screening in the full 2-flavor
QCD.
\section{FCM at finite temperatures}
The approach is based on the study of the quadratic
field correlators
$\,\,<tr\,F_{\mu\nu}(x)\Phi(x,0)F_{\lambda\sigma}(0)>$ ($x$ is
Euclidian), where $\Phi(x,0)$ is the Schwinger parallel
transporter
necessary to maintain gauge invariance. These correlators are
expressed in terms of two scalar functions, $D(x)$ and $D_1(x)$
\cite{NST:2009}. At $T\,=\,0$, the string tension $\sigma$ is
expressed only in terms of
$D(x)$:\begin{equation}\sigma\,=\,2\,\int\limits_0^{\infty}\,
d\lambda\,\int\limits_0^{\infty}\,d\nu\,D(\sqrt{\lambda^2+\nu^2})
.\end{equation} At $T\,\geq\,T_c$ one should distinguish between
electric and magnetic correlators $D^E(x)$, $D^H(x)$, $D^E_1(x)$,
and $D^H_1(x)$, and, correspondingly, between $\sigma^E$ and
$\sigma^H$. It was argued in \cite{Si:1991} and later confirmed on
the lattice \cite{lattice} that above the deconfinement region
$D^E(x)$ and $\sigma^E$ vanish, while the colorelectric correlator
$D^E_1(x)$ and colormagnetic correlators $D^H(x)$ and $D^H_1(x)$
should stay unchanged at least up to $T\sim 2\,T_c$. The
correlators $D^H(x)$ and $D^H_1(x)$ do not produce static
quark--antiquark potentials, they only define the spatial string
tension $\sigma_s\,=\,\sigma^H$ and the Debye mass
$m_D\propto\sqrt{\sigma_s}$ that grows with the temperature in the
dimensionally reduced limit \cite{AS:2006}.

The NP static $Q\overline Q$ potential at $T\,\geq\,T_c$
originates from the color--electric correlator function
$D^{E}_1(x)$\begin{equation}\label{eq:potential}
V_{np}(r,T)\,=\,\int\limits_0^{1/T}d\nu(1-\nu T)\int\limits_0^r
\lambda d\lambda\, D_1^{E}(x).\end{equation} In the confinement
region the function $D_1^{E}(x)$ was calculated in \cite{Si:2005}
exploiting the connection of field correlators to the  gluelump
 Green's function~\footnote{Recall that gluelumps are actually bound
states of the gluon field in a static color-octet source that have
been studied first in Lattice QCD \cite{12}.}
\begin{equation}\label{eq:DE} D^{E}_1(x)\,=\,{B}\,\,\frac{\exp(-M_0\,x)}{x},\end{equation}
where $ B=6\alpha_s^f\sigma_fM_0$, $\alpha_s^f$ being the freezing
value of the strong coupling constant, $\sigma_f$ is the sting
tension at $T=0$, and the parameter $M_0$ has the meaning of the
gluelump mass. Above $T_c$ the analytical form of $D_1^E$ should
stay unchanged at least up to $T\sim 2\,T_c$. The only change is
$B\to B(T)=\xi(T)B$, where the $T$-dependent factor
\begin{equation}\xi(T)\,=\,\left(1-0.36\,\,\frac{M_0}{B}\,\frac{T-T_c}{T_c}\right)\end{equation}
is determined by lattice data, see Eq. 52 of Ref.
\cite{DMSV:2007}.
Substituting (\ref{eq:DE}) into (\ref{eq:potential}) and
integrating over $\lambda$ one obtains
$V_{np}(r,T)\,=\,V(\infty,T)\,-\,V(r,T)$
where \begin{equation}\label{eq:vinf}
V(\infty,T)=\frac{B(T)}{M_0^2}\left[1-\frac{T}{M_0}\left(1-\exp\left(-\frac{M_0}{T}\right)\right)\right],
\end{equation} and \begin{equation}\label{eq:integral} V(r,T)=\frac{B(T)}{M_0}
\int\limits_0^{1/T}\, (1-\nu
T)e^{-\sqrt{\nu^2+r^2}\,M_0}\,d\nu.\end{equation} One observes the
characteristic feature of the static potential $V_{np}(r,T)$
produced by the correlator $D^E_1 (x)$: the potential  gives rise
to the constant term $V(\infty,T)$ in the $Q{\bar Q}$ interaction
at large distances, which can be viewed upon as the sum of
selfenergies of Q and ${\bar Q}$.\begin{table} \caption{Parameters
of the quark-antiquark potentials in units of GeV.
 Both sets of parameters correspond to $V(\infty,T_c)=0.505$ GeV} \centering\vspace{5mm}

\begin{tabular}{lrrr}\hline
$n_f$&$T_c$&$M_0$&$m_d(T_c)$\\\hline
0&0.275&0.9&0.793\\
2&0.165&1.08&0.545\\\hline
\end{tabular}

 \label{tab:parameters}\end{table}
\begin{table}
\caption{$J/\psi$ states above the deconfinement region. All the
quantities except for $T/T_c$ and $r_0$ are given in units of GeV,
the dimension of $r_0$ is GeV$^{-1}$.}
\begin{tabular}{ccccccc}\hline
$T/T_c$&~~$m_d$&~~{ $V(\infty,T)$}&$\mu_c$&{ $E_0(T)-V
(\infty,T)$}&$r_0$&$M_{c\overline{c}}$\\\hline
1&~~0.545&~~0.505&~~1.462&-\,0.026&6.89&3.281\\
1.2&~~0.609&~~0.433&~~1.438&-0.0098&11.45&3.334\\
1.3&~~0.640&~~0.398&~~1.426&-0.0046&16.86&3.164\\
1.4&~~0.671&~~0.365&~~1.415&-\,0.0013&25.51&3.164\\\hline
\end{tabular}
\label{tab:cc}\end{table}

\vspace{5mm}

\begin{table}
\caption{$\Upsilon$ states above the deconfinement region. The
notations are the same as in Table\ref{tab:cc}.}
\begin{tabular}{ccccccc}\hline
$T/T_c$&~~$m_d$&~~{ $V(\infty,T)$}&$\mu_b$&{ $E_0(T)-V
(\infty,T)$}&$r_0$&$M_{b\overline{b}}$\\\hline
1&~~0.545&~~0.504&~~4.948&-\,0.345&1.17&9.768\\
1.6&~~0.733&~~0.302&~~4.954&-\,0.182&1.54&9.725\\
2.0&~~0.853&~~0.187&~~4.937&-\,0.102&2.01&9.688\\
2.8&~~1.082&~~0&~~4.837&-\,0.008&7.88&9.592
\\\hline
\end{tabular}
\label{tab:bb}\end{table}
\begin{table}\caption{Dissociation temperatures (in units of $T_c$) for
 $c\overline{c}$,
 $b\overline{b}$, and $\Omega_{bbb}$ states. $\Omega_{ccc}$ is
 unbound both for $n_f=0$ and $n_f=2$.}

\begin{tabular}{ccccc}\hline
&$J/\psi$&$\Upsilon$&$\Omega_{bbb}$\\\hline
$n_f=0$&~~1.29&2.57&1.8\\
$n_f=2$&~~1.48&2.96&2.35\\
\hline
\end{tabular}
 \vspace{3mm}
\label{tab:melting}\end{table}

\begin{table}
\caption{$\Omega_{bbb}$ state above the deconfinement region. The
interquark distances
$\sqrt{<\,r_{ij}^2\,>}\,=\,{\sqrt\frac{<R^2>}{\mu_b}}$.}
\vspace{4mm}

\begin{tabular}{ccccccc}\hline
$T/T_c$&~~$m_d$&~~{ ${\cal V}(\infty,T)$}&$\mu_b$&{ $E_0(T)-{\cal
V} (\infty,T)$}&$\sqrt{<R^2>}$&$M_{bbb}$\\\hline
1&~~0.545&~~0.757&~~4.962&-\,0.327&3.44&14.837\\
1.4&~~0.672&~~0.548&~~4.926&-\,0.185&~4.22&14.768\\
2.0&~~0.853&~~0.281&~~4.919&-0.192&~7.56&14.641\\
2.3&~~0.940&~~0.166&~~4.830&-\,0.0034&~18.50&14.563\\
2.4&~~0.969&~~0.131&~~4.812&+\,0.0021&~32.40&14.533
\\\hline
\end{tabular}
\label{tab:bbb}\end{table}

%

We can now exploit the relativistic Hamiltonian technics
\cite{DKS} successfully applied for mesons, baryons, glueballs and
hybrids in the confinement phase. This technic does not take into
account chiral degrees of freedom and is applicable when
spin-dependent interaction can be treated as perturbation.
Therefore below we consider only heavy quarkonia and heavy
baryons, leaving light quarkonia with chiral symmetry restoration
to another publication.

Recall that, in the framework of the FCM, the masses of heavy
quarkonia are defined as \begin{equation} \label{eq:mass}M_{Q\bar
Q}\,=\,\frac{m_Q^2}{\mu_{Q}}\,+\,\mu_Q\,+\,E_0(m_Q,\mu_Q),\end{equation}
$E_0(m_Q,\mu_Q)$ is an eigenvalue of the Hamiltonian
\begin{equation}\label{H}H=H_0+V_{np}+V_{OGE},\end{equation}where we
have omitted spin-dependent and self-energy terms proportional to
$1/\mu_Q$. In Eq. (\ref{H}) $V_{OGE}$ is the one-gluon-exchange
potential which is expected to be exponentially screened at large
distances
\begin{equation}\label{eq:alpha}V_{OGE}(r,T)\,=\,-\,\frac{4}{3}\,\frac{\alpha_s}{r}
\exp(-m_d(T)\,r),\end{equation}
$m_d(T)$ being the Debye mass. In Eq. (\ref{eq:mass}) $m_Q$ are
the bare quark masses, and einbeins $\mu_Q$ are treated as
c-number variational parameters~\footnote{The eigenvalues
$E_0(m_Q,\mu_Q)$ are found as functions of the bare quark masses
$m_Q$ and einbeins $\mu_Q$, and are finally minimized with respect
to the $\mu_Q$. Once $m_Q$ is fixed, the quarkonia spectrum is
described.}. With such simplifying assumptions the spinless
Hamiltonian takes an apparently nonrelativistic form, with einbein
fields playing the role of the constituent masses of the quarks.
In what follow we take $m_c=1.4$ GeV, $m_b=4.8$ GeV. As in the
confinement region, the constituent masses $\mu_Q$ only slightly
exceed the bare quark masses $m_Q$ that reflect smallness of the
kinetic energies of heavy quarks. The dissociation points are
defined as those temperature values for which the energy gap
between $V(\infty,T)$ and $E_0$ disappears.

\section{Quark-antiquark states}

The non-perturbative quark-antiquark potential is defined by the
two parameters $B$ and $M_0$. In what follows we take
$\sigma_f=0.18$ GeV$^2$ and $\alpha_s^f=0.6$, so that
$B\,=\,0.648\,M_0$, and vary the gluelump mass $M_0$  around the
central value $M_0=1$ GeV in order to maintain  the asymptotic
value $V(\infty,T_c)=0.505$ GeV. This value agrees with lattice
estimate for the free quark-antiquark energy~\footnote{~However,
the difference in the parameter $M_0$ causes the small difference
of $V(\infty,T)$ for $T\,>\,T_c$.}. The strong coupling constant
was taken $\alpha_s=0.35$~\footnote{The account of the running
$\alpha_s(r)$ in the Coulomb potential produces a tiny effect as
compared with the case of a constant $\alpha_s=0.35$ both for the
energies
and wave functions \cite{NSV:2011}.}. The parameters of the
potential
 are listed in Table \ref{tab:parameters}, where for the
reference we also indicate the values of the Debye mass $m_d(T_c)$
\cite{A:2003}.

Some details of our calculation for the full $n_f\,=\,2$ QCD can
be inferred from Tables \ref{tab:cc}, \ref{tab:bb}~\footnote{The
corresponding results for the pure gluodynamics ($n_f\,=\,0$) are
given in Ref. \cite{NSV:2011}}. At $T=T_c$ we obtain the weakly
bound $c{\overline c}$ state. The melting temperature is
$\sim\,1.3\,T_c$ for $n_f\,=\,0$ and $1.48\,T_c$ for $n_f\,=\,2$,
see Table \ref{tab:melting}. The charmonium masses lie in the
interval 3.1 - 3.3 GeV.
Note that at the melting point
$r(J/\psi)\to\infty$ that is consistent with nearly-free dynamics.

As expected, the $\Upsilon$ state is much more bound and remains
intact up to the larger temperatures, $T\,\sim\, 2.3\,T_c$. This
is in agreement with the lattice study of Ref. \cite{aarts:2010}.
The masses of the L = 0 bottomonium lie in the interval 9.6--9.8
GeV, about 0.2--0.3 GeV higher than 9.460 GeV, the mass of
$\Upsilon(1S)$ at $T=0$. At $T=T_c$ the $b{\overline b}$
separation $r_0$ is 0.25 fm that is compatible with $r_0=0.28$ fm
at $T=0$ (at the melting point $r_0\to\infty$). Note that the 1S
bottomoniium undergos very little modification till
 $T\,\sim 2\,T_c$.
The melting temperatures for the
$J/\psi$ and $\Upsilon$ are shown in Table \ref{tab:melting}.

\section{$QQQ$ baryons}\label{section:QQQ}
The three quark potential is given by
$V_{QQQ}\,=\,\frac{1}{2}\,\sum_{i<j}\,V_{Q{\bar Q}}(r_{ij},T)$,
where $\frac{1}{2}$ is the color factor and  $V_{Q{\bar Q}}$ is
the sum of the perturbative and NP quark-antiquark potential. We
solve the three quark Schr\"{o}dinger equation by the
hyperspherical harmonics method. The wave function
in the
hypercentral approximation is written as
\begin{equation}
\Psi(R,T)\,=\,\frac{1}{\sqrt{\pi^3}}\,\frac{u(R,T)}{R^{5/2}},
\end{equation}
where the hyperradius
\begin{equation}R^2\,=\,
\frac{\mu_Q}{3}\,\left(r_{12}^2\,+\,r_{23}^2\,+\,r_{31}^2\right)
\end{equation}
is invariant under quark permutations. Averaging the three--quark
potential over the six-dimensional sphere one obtains the
one-dimensional Schr\"odinger equation for the reduced function
$u(R,T)$ \beqn\label{eq:se}\frac{d^2
u(R,T)}{dR^2}+2\left[E_0-\frac{15}{8\,R^2}-\frac{3}{2}{\cal
V}(R,T)\right]u(R,T)=0,\eeqn where ${\cal V}(R,T)={\cal
V}_{OGE}(R,T)+{\cal V}_{np}(R,T)$ and
\begin{equation} {\cal V}_{OGE}(R,T)=
-\frac{16\,\alpha_s}{3\,\pi}\,\int\limits_0^{\pi/2}\frac{
\exp(\,-\,m_d(T)\hat R)}{\hat R}\,\sin^2(2\theta),\end{equation}
\beqn\label{eq:v3Q}&& {\cal
V}_{np}(R,T)=V(\infty,T)-\frac{4\,\xi(T)B}{\pi
M_0}\times\nonumber\\&&
\int\limits_0^{\pi/2}\left({\hat R}K_1({\hat
R})-\frac{T}{M_0}\,e^{-{\hat R}}(1+{\hat
R})\right)\sin^2(2\theta)\,d\theta,\,\,\,\,\,\,\,\,\,\,\,\,\,\,\,\,\,\,\,\,\,\eeqn
$V(\infty,T)$ being given by Eq. (\ref{eq:vinf}), and ${\hat
R}=2\,M_0\,R\sin\theta/\mu_Q$. In Eq. (\ref{eq:v3Q}) we use the
approximate expression for the non-perturbative $Q{\bar Q}$
potential (\ref{eq:integral})
\begin{equation}\label{eq:approximation} V(r,T)\approx
\frac{B(T)}{M_0^2}\left(xK_1(x)\,-\,\frac{T}{M_0}\exp(-x)(1+x)\right),\end{equation}
where $x=M_0r$ and $K_1(x)$ is the McDonnald function,

The temperature dependent mass of the colorless $QQQ$ states is
\begin{equation}\label{eq:mQQQ} M_{QQ
Q}\,=\,\frac{3}{2}\frac{m_Q^2}{\mu_{Q}}\,+\,\frac{3}{2}\,\mu_Q\,+\,E_0(m_Q,\mu_Q),\end{equation}
where $\mu_Q$ are now defined from the extremum condition imposed
on $M_{QQQ}$ in (\ref{eq:mQQQ}).
The bound $QQQ$ state
exists if $E_0(m_Q,\mu_Q)\,\leq\,{\cal V}_{QQQ}(\infty,T)$, where
\begin{equation}\label{eq:VQQQinfinity}{\cal
V}_{QQQ}(\infty,T)\,=\,\frac{3}{2}\,V(\infty,T).\end{equation}
 There is no bound $\Omega_{ccc}$
states~\footnote{However, in all our calculations the
$\Omega_{ccc}$ was found to lie almost at threshold. For example
for $n_f=2$ we obtain $E_0(T_c)-V(\infty,T_c)\,=\,+1.2$ MeV.} but
the $\Omega_{bbb}$ survives up to $T\sim 1.8-2.4\,\,T_c$
(depending on $n_f$), see Tables
 \ref{tab:melting}, \ref{tab:bbb}.

\section{Conclusions}
The static $Q{\bar Q}$ potential has been extensively investigated
within the FCM and provides useful tool to study in-medium
modification of inter-quark forces. This potential
provides also useful  quantitative insights into the problem of
quarkonium binding in QGP. In particular, the color electric
forces due to the nonconfining correlator $ D^E _1$ survive in the
deconfined phase and can support bound states at $T\,>\,T_c$. In
this paper, we used a FCM approach to the problem of heavy quark
potentials at finite temperature. We have calculated binding
energies and melting temperatures for the lowest eigenstates in
the $c\overline{c}$, $b\overline{b}$, and $bbb$ systems neglecting
spin-dependent and self-energy terms in the Hamiltonian.
We find that the ground state of $J/\psi$ survives up to
$T\,\sim\,1.3-1.5\,T_c$, and there is no bound $\Omega_{ccc}$
state at $T\geq\,T_c$. On the other hand, the $b\overline{b}$ and
$bbb$ states survive up to higher temperature, $T\sim
2.6-3.0\,T_c$ and $T\sim 1.8-2.4\,T_c$ for $n_f=0,2$,
respectively. The results suggest that the systems are strongly
interacting above $T_c$.

\begin{theacknowledgments}
\noindent This work was supported in part by RFBR grants
08-02-00657, 08-02-00677, and
 09-02-00629.

\end{theacknowledgments}




\bibliographystyle{aipproc}

\begin{thebibliography}{9}
\bibitem{T:2009} M.~J.~Tannenbaum, {\it Rep.\ Prog.\ Phys.}
\textbf{69},  2005 (2006).
\bibitem{Linde:1980} A.~D.~Linde, {\it Phys.\ Lett.} B\textbf {96}, 289 (1980).
\bibitem{[16]} O.~Kaczmarek, F.~Karsch,
P.~Petreczky and F.~Zantow, {\it Nucl.\ Phys.\ Proc.\ Suppl.}
\textbf{129}, 560 (2004); M.~D\"{o}ring, S.~Ejiri, O.~Kaczmarek,
F.~Karsch, E.~Laermann, hep-lat~/~0509150.

\bibitem{NST:2009}
A.~V.~Nefediev, Yu.~A.~Simonov, M.~A.~Trusov, {\it Int.\ J.\ Mod.\
Phys.} \textbf{E18}, 549 (2009).
\bibitem{NSV:2009}I.~M.~Narodetskiy, Yu.~A.~Simonov,
A.~I.~Veselov, {\it JETP\ Lett.} \textbf{90}, 232 (2009).
\bibitem{Si:1991} Yu.~A.~Simonov, {\it JETP\ Lett.} \textbf{54}, 249
(1991), {\it Phys.\ Atom.\ Nucl.} \textbf{58}, 309 (1995).
\bibitem{lattice} M.~D'Elia, A.~Di Giacomo, and E.~Meggiolaro, {\it Phys.\ Rev.}
D\textbf{67}, 114504 (2003); G~.S.~Bali, N.~Brambilla, A.~Vairo,
{\it Phys.\ Lett.} B\textbf{421}, 265 (1998).
\bibitem{AS:2006} N.~O.~Agasian and Yu.~A.~Simonov, {\it Phys.\ Lett.} B\textbf{639}, 82
(2006).
\bibitem{Si:2005} Yu.~A.~Simonov, {\it Phys.\ Lett.} B\textbf{619}, 293
(2005).
\bibitem{12}
M.~ Foster and C.~Michael, {\it Phys.\ Rev.} D\textbf
{59},~094509~(1999); G.~S.~Bali and A.~Pineda, {\it Phys.\ Rev.}
D\textbf{69},~094001~ (2004).
\bibitem{DMSV:2007}
A.~DiGiacomo, E.~Meggiolaro, Yu.~A.~Simonov, and A.~I.~ Veselov,
{\it Phys.\ Atom.\ Nucl.} \textbf{70}, 908 (2007).

\bibitem{A:2003} N.~O.~Agasian, {\it Phys.\ Lett.} B\textbf{562},
~257 (2003). \bibitem{DKS} A.~Yu.~Dubin, A.~B.~Kaidalov, and
Yu.~A.~Simonov, {\it Phys.\ Lett.} B\textbf{323}, 4 (1994); {\it
Phys.\ Atom.\ Nucl.} \textbf {56}, 1745 (1993).

\bibitem{NSV:2011} I.~M.~Narodetskiy, Yu.~A.~Simonov, and
A.~I.~Veselov, {\it Phys.\ Atom.\ Nucl.} \textbf{74}, No 3 (2011),
in print

\bibitem{aarts:2010} G. Aarts {\it et al.},
arXiv: hep-lat~/~1010.3725


\end{thebibliography}

\end{document}